\documentclass[11pt]{article}
\usepackage{latexsym}
\usepackage{amssymb}
\usepackage{epsf}
\usepackage{amsmath}
\usepackage[hypertex]{hyperref}
\usepackage{graphicx}% Include figure files

\usepackage{slashed}

\begin{document}
\pagestyle{empty}

\begin{flushright}
TU-890
\end{flushright}

\vspace{3cm}

\begin{center}

{\bf\LARGE Hidden local symmetry
and color confinement}
\\

\vspace*{1.5cm}
{\large 
Ryuichiro Kitano
} \\
\vspace*{0.5cm}

{\it Department of Physics, Tohoku University, Sendai 980-8578,
 Japan}\\

\end{center}

\vspace*{1.0cm}

\begin{abstract}
{\normalsize
The hidden local symmetry is a successful model to describe the
 properties of the vector mesons in QCD. We point out that if we
 identify this hidden gauge theory as the magnetic picture of QCD, a
 linearized version of the model simultaneously describes color
 confinement and chiral symmetry breaking.
We demonstrate that such a structure can be seen in the Seiberg dual
 picture of a softly broken supersymmetric QCD. The model possesses
 exact chiral symmetry and reduces to QCD when mass parameters are taken
 to be large.
Working in the regime of the small mass parameters, we show that there
 is a vacuum where chiral symmetry is spontaneously broken and
 simultaneously the magnetic gauge group is Higgsed. If the vacuum we
 find persists in the limit of large mass parameters, one can identify
 the $\rho$ meson as the massive magnetic gauge boson, that is an
 essential ingredient for color confinement.
}
\end{abstract} 

%%%%%%%%%%%%%%%%%%%%%%%%%%%%%%%%%%%%%%%%%%%%%%%%%%%%%%%%%%%%%%%%%%%%%%%%%%%%
\newpage
\baselineskip=18pt
\setcounter{page}{2}
\pagestyle{plain}
\baselineskip=18pt
\pagestyle{plain}

\setcounter{footnote}{0}

\section{Introduction}

In order to understand the mechanisms for color confinement and chiral
symmetry breaking in QCD, one probably needs a non-perturbative
method. A hint for this may be the hidden local
symmetry~\cite{Bando:1984ej}; the properties of the $\rho$ meson are
well described as the gauge boson of a spontaneously broken gauge
symmetry. It is suggesting that there is some dual picture of QCD in
which the $\rho$ meson comes out as a gauge field.

One of such examples is the holographic QCD~\cite{Sakai:2004cn,
Erlich:2005qh, Da Rold:2005zs}. The $\rho$ meson in this case is
identified as the lowest Kaluza-Klein mode of a gauge field which
propagates into the extra-dimension in the gravity picture. In order to
fully connect to the real 4D QCD, one needs to decouple all the
artificially added modes. If one can do this decoupling smoothly, {\it
i.e.,} without a phase transition, qualitative understanding of the low
energy QCD can be obtained within the perturbation theory in the dual
picture.

Another example, which is more relevant to this work, is the use of the
electric-magnetic dualities in supersymmetric gauge
theories\footnote{See, for example, Ref.~\cite{Strassler:1997ny} and
references therein for a review on the supersymmetric gauge theories,
dualities and their connection to real QCD.}.
Seiberg and Witten~\cite{Seiberg:1994rs, Seiberg:1994aj} have shown that
${\cal N}=2$ supersymmetric $SU(2)$ gauge theory provides an explicit
example of confinement by the monopole
condensation~\cite{'tHooft:1981ht, Mandelstam:1974vf}.
Along this line, relations between flavor symmetry breaking and
confinement have been studied. It has been observed that there are
examples in which flavor symmetry is spontaneously broken by
condensations of magnetic degrees of freedom~\cite{Carlino:2000ff,
Konishi:2005qt, Eto:2006dx, Gorsky:2007ip}, and thus the
same condensations explain both confinement and flavor symmetry
breaking.
In the case of ${\cal N}=1$ supersymmetric QCD (SQCD), the Seiberg
duality~\cite{Seiberg:1994pq} is believed to be the electric-magnetic
duality. The dual magnetic theory contains dual quarks, which transform
under both the dual gauge group and the flavor group. The condensation
of the dual quarks causes the color-flavor locking in the dual picture,
and again relates confinement and flavor symmetry breaking. This
phenomenon has been demonstrated in a model with a $U(N)$ gauge
group~\cite{Shifman:2007kd}.
Recently, Komargodski discussed the identification of massive magnetic
gauge bosons in the Seiberg dual picture as the vector mesons with
emphasis on the similarity to the hidden local symmetry and the
realization of the vector meson dominance~\cite{Komargodski:2010mc}.
See also Ref.~\cite{Harada:1999zj} for an early work on this
interpretation.

The examples studied in supersymmetric theories are very suggestive that
real QCD may have a similar picture for confinement and chiral symmetry
breaking. However, there are difficulties in connecting non-perturbative
results in supersymmetric theories to real QCD. First of all, real QCD
has no supersymmetry. Even with supersymmetry, $SU(3)$ gauge theories
with two or three flavors are out of the region where the Seiberg
duality exists. It is also not easy to have chiral symmetry which is an
essential feature in QCD. If we start with ${\cal N} = 2$ supersymmetric
theory with $SU(3)$ gauge group, chiral symmetry is explicitly broken
from the first place. One can start with ${\cal N}=1$ theories but no
explicit example with exact chiral symmetry and its breaking to their
diagonal subgroup has been found in Refs.~\cite{Shifman:2007kd,
Komargodski:2010mc}.

As for the problem of supersymmetry, Aharony {\it et al.}~studied SQCD
models with small soft supersymmetry breaking
terms~\cite{Aharony:1995zh}. These models reduce to QCD when the soft
terms are large.
Many interesting results are obtained by using the exact results of
SQCD~\cite{Seiberg:1994bz, Seiberg:1994pq}. For example, in $SU(N_c)$
gauge theory with $N_f$ massless flavors and for $N_f < N_c$,
spontaneous chiral symmetry breaking by the meson condensation can be
seen as a result of balancing between the non-perturbatively generated
superpotential and the soft masses.
This vacua has been further studied in Ref.~\cite{Martin:1998yr}, and
the different scaling behavior from QCD expectations in the large $N_c$
limit has been reported.
So far, it is not conclusive if the vacuum with the meson condensation
continuously connects to the QCD vacuum in the limit of large
supersymmetry breaking terms.

In this paper, we extend the work of Ref.~\cite{Aharony:1995zh} and
consider the realization of the hidden local symmetry as the Seiberg
dual gauge theory~\cite{Komargodski:2010mc}. Our starting point is
similar to the model in Ref.~\cite{Shifman:2007kd} where extra $N_c$
flavors are introduced in addition to the massless $N_f$ quarks in SQCD
with the $SU(N_c)$ gauge group. The extra $N_c$ flavors are massive and
play a role of the regulator, in contrast to the model in
Ref.~\cite{Shifman:2007kd} where masses are added to $N_f$ flavors. Our
model has exact chiral symmetry, $SU(N_f)_L$ $\times$ $SU(N_f)_R$, as
well as the $U(1)_B$ baryon symmetry.
The Seiberg dual theory is an $SU(N_f)$ gauge theory and contains dual
quarks which transform under both the gauge group and the flavor group; 
this part has the same structure as the hidden local symmetry.
By including supersymmetry breaking terms, we find a vacuum where the
condensations of the dual quarks break chiral symmetry down to the
vectorial $SU(N_f)_V$ symmetry while preserving $U(1)_B$ symmetry as in
real QCD.
The Higgsing of the dual magnetic gauge theory expels the magnetic flux
(that is the color flux in the electric picture) from the vacuum, and
thus two phenomena in QCD: confinement and chiral symmetry breaking, are
connected.
The model reduces to QCD when we take a limit of large masses of the
auxiliary flavors and large soft supersymmetry breaking terms. We argue
that there is a good chance that the vacuum we find is continuously
connected to the QCD vacuum since they are pretty similar.
The vacuum is different from the one discovered in
Ref.~\cite{Aharony:1995zh}. Adding extra flavors is essential for our
vacuum to exist.
The non-perturbatively generated superpotential is not necessary for
stabilizing the vacuum at non-zero vacuum expectation values
(VEVs). Therefore, there is no problem in the large $N_c$ scalings.

We start with the review of the hidden local symmetry and comment on the
vector meson dominance in linearized models. We find that the vector
meson dominance cannot be realized in SQCD models at tree level when the
symmetry breaking pattern is $SU(N_f)_L$ $\times$ $SU(N_f)_R$ $\to$
$SU(N_f)_V$, that has not been studied in
Ref.~\cite{Komargodski:2010mc}.
We present an explicit model to realize hidden local symmetry in
Section~\ref{sec:model} and study the vacuum. The case with $N_c = 3$
and $N_f = 2$ is studied in Section~\ref{sec:qcd} where we propose a new
interpretation of the light mesons. An application to electroweak
symmetry breaking is discussed in Section~\ref{sec:ewsb}.

\section{Hidden local symmetry and vector meson dominance}

We start with the discussion of the hidden local symmetry and its
connection to the vector meson dominance. We discuss the difficulty in
realizing the vector meson dominance at tree level in the SQCD-like
models.

\subsection{Hidden local symmetry}

The hidden local symmetry is a model for the $\rho$ meson and the pions
based on a spontaneously broken $SU(N_f)_{\rm local}$ gauge
symmetry. The chiral symmetry $SU(N_f)_L$ $\times$ $SU(N_f)_R$ is
spontaneously broken (non-linearly realized), and the breaking
simultaneously gives a mass for the gauge boson, which is identified as
the $\rho$ meson.
The Lagrangian is given by
\begin{eqnarray}
 {\cal L} &=& - {1 \over 4 g_H^2} F^{a}_{\mu \nu} F^{a \mu \nu}
\nonumber \\
&& + {a f_\pi^2 \over 2} {\rm tr} \left[
|D_\mu U_L |^2 + |D_\mu U_R|^2
\right]
\nonumber \\
&& + {(1-a) f_\pi^2 \over 4} {\rm tr} \left[
|\partial_\mu (U_L U_R) |^2
\right].
\label{eq:hls}
\end{eqnarray}
The fields $U_L$ and $U_R$ are unitary matrices, and transform as
\begin{eqnarray}
 U_L \to g_L U_L g^{-1}, \ \ \ 
 U_R \to g U_R g_R^{-1},
\end{eqnarray}
under group elements,
\begin{eqnarray}
 g_L \in SU(N_f)_L, \ \ \ g \in SU(N_f)_{\rm local}, \ \ \ g_R \in SU(N_f)_R,
\end{eqnarray}
where $SU(N_f)_{\rm local}$ is gauged. The covariant derivatives are
\begin{eqnarray}
 D_\mu U_L = \partial_\mu U_L + i U_L A^a_\mu T^a,
\end{eqnarray}
\begin{eqnarray}
 D_\mu U_R = \partial_\mu U_R - i A^a_\mu T^a U_R.
\end{eqnarray}

\begin{figure}[t]
\begin{center}
 \includegraphics[width=8cm]{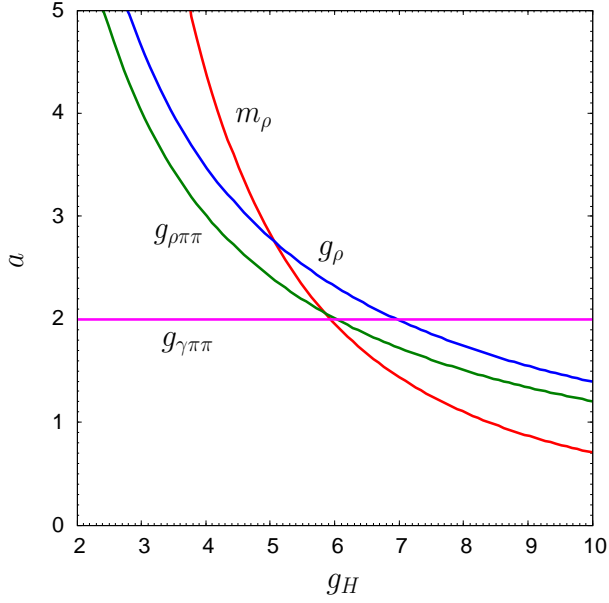} 
\end{center}
\caption{The predictions of the hidden local symmetry. We have used 
$m_\rho = 776~{\rm MeV}$, $f_\pi = 92.4~{\rm MeV}$, $g_{\rho \pi \pi} =
 6.03$,
$g_\rho = (345~{\rm MeV})^2$, $g_{\gamma \pi \pi} \sim 0$.
Values are taken from Ref.~\cite{Erlich:2005qh}.
}
\label{fig:HLS}
\end{figure}

In the unitary gauge, $U_L = U_R$, the massless pion $\pi^a$ is embedded
as
\begin{eqnarray}
 U_L = U_R = e^{i \pi^a T^a / f_\pi^2}.
\end{eqnarray}
The gauge boson $A_\mu^a$ obtains a mass from the kinetic terms of $U_L$
and $U_R$. The massive gauge boson describes the $\rho$ meson.

This Lagrangian gives phenomenologically successful nontrivial relations
among physical quantities,
\begin{eqnarray}
 m_\rho^2 = a g_H^2 f_\pi^2,
\end{eqnarray}
\begin{eqnarray}
 g_{\rho \pi \pi} = {a \over 2} g_H,
\end{eqnarray}
\begin{eqnarray}
 g_{\gamma \pi \pi} = - {a-2 \over 2} e,
\end{eqnarray}
\begin{eqnarray}
 g_\rho = a g_H f_\pi^2.
\end{eqnarray}
These relations are quite successful with
\begin{eqnarray}
 g_H \simeq 6,\ \ \  a \simeq 2.
\end{eqnarray}
The fact that $g_{\gamma \pi \pi}$ vanishes for $a=2$ is called the
vector meson dominance and realized in QCD. Fig.~\ref{fig:HLS}
demonstrates how successful the model is.

\subsection{Linearized hidden local symmetry and the value of $a$}

The hidden local symmetry is a non-linear sigma model due to the
constraint that $U_L$ and $U_R$ are unitary matrices. The model can
easily be UV completed by embedding $U_L$ and $U_R$ into some linearly
transforming Higgs fields.

For example, $U_L$ and $U_R$ can be embedded into $q: (N_f, \overline
{N_f}, 1)$ and $\bar q: (1, N_f, \overline {N_f})$, respectively. The
Lagrangian (the kinetic terms for the Higgs fields) reduces to the
hidden local symmetry with $a=1$ at tree level.
When one obtains the hidden local symmetry from deconstruction of the
extra dimensional gauge theory, this value is realized at the three-site
level~\cite{Belyaev:2009ve}, and becomes $a=4/3$ in the continuum
limit~\cite{Da Rold:2005zs, Belyaev:2009ve}.

Another example is to include $M: (N_f, 1, \overline {N_f})$ in addition
to the above model, and embed $U_L U_R$ to $M$. This reduces to a model
with $0 < a \leq 1$. The correct sign for the kinetic term of $M$
indicates that $a>1$ is not possible at tree level as one can see in
Eq.~\eqref{eq:hls}.
When we identify $SU(N_f)_{\rm local}$ as the magnetic gauge theory of
SQCD, the Seiberg duality says that the particle content is $q$, $\bar
q$ and $M$ as the dual quarks and the meson. Therefore, one cannot
obtain $a>1$ in the dual picture of the SQCD at tree level.
In Ref.~\cite{Komargodski:2010mc}, it is argued that $a = 2$ is realized
in SQCD. However, the examples used there is not the chiral symmetry
breaking.

Although it sounds unfortunate that the phenomenologically favorable
value, $a \sim 2$, cannot be realized in SQCD, we do not argue that the
approach from SQCD is unsuccessful.
It is important to note that the value $a=2$ is not stable under the
renormalization. In Ref.~\cite{Harada:1992bu}, it was found that the $a$
parameter has a UV fixed point at $a=1$. Therefore, having $a \sim 1$ at
tree level may not be so bad.

\section{An SQCD model as a regularization of QCD}
\label{sec:model}

We present a model to study QCD by adding massive modes as
regulators. Although it is not guaranteed that the study of such models
have something to do with the real QCD, it provides us with qualitative
understanding of it if it is smoothly connected to the
vacuum of real QCD. We present a model which possesses a sufficiently
similar vacuum to the real QCD one.

\subsection{Set up}
Our goal is to study a non-supersymmetric $SU(N_c)$ gauge theory with
$N_f$ massless quarks. We here add various massive modes to make it
possible to use the Seiberg duality.

The starting point is an ${\cal N}=1$ supersymmetric $SU(N_c)$ gauge
theory with $N_f + N_c$ chiral superfields in the fundamental
representation.  We list the particle content and the quantum numbers in
Table~\ref{tab:ele}.
Extra $N_c$ quarks are added so that the dual gauge group is $SU(N_f)$.

\renewcommand{\arraystretch}{1.3}
\begin{table}[t]
\begin{center}
\small
 \begin{tabular}[t]{cccccccc}
& $SU(N_c)$ & $SU(N_f)_L$ & $SU(N_f)_R$ & $U(1)_B$ 
& $SU(N_c)_V$ &  $U(1)_{B^\prime}$ & $U(1)_R$ 
\\ \hline \hline
 $Q$ & $N_c$ & $N_f$ & $1$ & 1 & 1 & 0 &
$(N_f - N_c)/N_f$ \\
 $\overline Q$ & $\overline {N_c}$& 1 & $\overline {N_f}$ & $-1$ 
& 1 & 0 & $(N_f - N_c)/N_f$ \\ \hline
 $Q^\prime$ & $N_c$ & 1 & $1$ & 0 & $\overline{N_c}$ & 1 &
1 \\
 $\overline Q^\prime$ & $\overline {N_c}$& 1 & 1 & 0 
& ${N_c}$ & $-1$ & 1 \\ 
 \end{tabular}
\end{center}
\caption{Quantum numbers in the electric picture.}
\label{tab:ele}
\end{table}
\renewcommand{\arraystretch}{1}

In order to remove unwanted modes, first we add a mass term for the
extra quarks,
\begin{eqnarray}
 W = m Q^\prime \bar Q^\prime.
\label{eq:mass}
\end{eqnarray}
We also gauge the artificially enhanced $U(1)_{B^\prime}$ symmetry so
that the breaking of it would not give the Goldstone mode.
Finally, we add soft supersymmetry breaking terms to reduce the model to
the non-supersymmetric QCD,
\begin{eqnarray}
 {\cal L}_{\rm soft}  = 
- \tilde m^2 (|Q|^2 + |\bar Q|^2 + |Q^\prime|^2 + |\bar Q^\prime |^2)
- \left(
{m_\lambda \over 2} \lambda \lambda
+ {\rm h.c.}\right)
- 
\left(
B m Q^\prime \bar Q^\prime
+ {\rm h.c.}\right)
,
\end{eqnarray}
where the first and the second terms are the scalar and the gaugino
masses, respectively. The last term is the $B$-term associated with the
mass term in Eq.~\eqref{eq:mass}.

For $N_f + N_c \leq 3N_c/2$, {\it i.e.,} $N_f \leq N_c/2$, the dual
magnetic picture is a free theory in the IR and thus analysis in the
perturbation theory is possible. The use of the Seiberg duality and the
perturbative expansions are justified when the mass parameters $m$,
$\tilde m$, $m_\lambda$, and $B$ are all smaller than the dynamical
scale.
For $N_c/2 < N_f < 2 N_c$, the theory is in the conformal window. The
dual description is more weakly coupled for $N_f < N_c$.

\subsection{Magnetic description}

\renewcommand{\arraystretch}{1.3}
\begin{table}[t]
\begin{center}
\hspace*{-.8cm}
\small
 \begin{tabular}[t]{ccccccccc}

& $SU(N_f)$ & $SU(N_f)_L$ & $SU(N_f)_R$ & $U(1)_B$ 
& $SU(N_c)_V$  & $U(1)_{B^\prime}$ & $U(1)_R$ 
\\ \hline \hline
 $q$ & $N_f$ & $\overline{N_f}$ & 1 & 0 
& 1 & $N_c/N_f$ &  $N_c/N_f$ \\ 
 $\overline q$ & $\overline {N_f}$ & 1 & $N_f$ & 0 
& 1 & $-N_c/N_f$ &  $N_c/N_f$ \\ 
 $\Phi$ & 1 & $N_f$ & $\overline{N_f}$ & 0 
& 1 & 0 &  $2 (N_f - N_c)/N_f$ \\ \hline
 $q^\prime$ & $N_f$ & 1 & 1 & 1 
& ${{N_c}}$ & $-(N_f - N_c)/N_f$ &  0 \\ 
 $\overline q^\prime$ & $\overline {N_f}$ & 1 & 1 & $-1$ 
& $\overline{N_c}$ & $(N_f - N_c)/N_f$ &  0 \\ 
 $Y$ & 1 & 1 & $1$ & 0 & 1 + Adj. & 0 &
2 \\
 $Z$  & 1 & 1 & $\overline {N_f}$ & $-1$ 
& $\overline{N_c}$ & 1 & $(2 N_f - N_c)/N_f$ \\ 
 $\overline Z$ & 1 & $N_f$ & 1 & 1 & ${N_c}$ & $-1$ &
$(2 N_f - N_c)/N_f$ \\

 \end{tabular}
\end{center}
\caption{Quantum numbers in the magnetic picture.}
\label{tab:mag}
\end{table}
\renewcommand{\arraystretch}{1}

The dual picture is an $SU(N_f)$ gauge theory as is designed. We would
like to identify this gauge theory as the hidden local symmetry.
The degrees of freedom in the dual description are mesons $M$ and the
dual quarks: $q$, $\bar q$, $q^\prime$, $\bar q^\prime$. 
The quantum
numbers are listed in Table~\ref{tab:mag}, where we decomposed the meson
as follows:
\begin{eqnarray}
 M = \left(
\begin{array}{cc}
 Y & Z \\
 \bar Z & \Phi \\
\end{array}
\right).
\end{eqnarray}
The superpotential is given by
\begin{eqnarray}
 W = m \Lambda Y + h \left(
q^\prime Y \bar q^\prime 
+ q^\prime Z \bar q + q \bar Z \bar q^\prime 
+ q \Phi \bar q
\right),
\label{eq:superDual}
\end{eqnarray}
where $\Lambda$ is a parameter of the order of the dynamical scale, and
$h$ is a dimensionless coupling constant.

At the supersymmetric level, as is well known, the potential has a
runaway direction after including the non-perturbative effects, $\langle
Y \rangle \neq 0$ and $\langle \Phi \rangle \to \infty$. The dual gauge
theory is confined (electric gauge theory is Higgsed). Stabilizing this
direction by supersymmetry breaking terms, one can find a vacuum studied
in Ref.~\cite{Aharony:1995zh}.

As another possibility, $\langle q^\prime \bar q^\prime \rangle \neq 0$
looks minimizing the potential at tree level. Because the rank of
$q^\prime \bar q^\prime $ is smaller than $N_c$, there is not a stable
supersymmetric vacuum in this direction, but there can be a local
minimum as in the model of Ref.~\cite{Intriligator:2006dd}.
The possibility of such a local minimum has been studied in the presence
of massless flavors and found that there is
not~\cite{Giveon:2008wp}. However, adding supersymmetry breaking terms
may create a vacuum. In this case, the dual gauge boson obtains a mass
by the $q^\prime \bar q^\prime$ condensation without breaking the
$SU(N_f)_L$ $\times$ $SU(N_f)_R$ chiral symmetry. The structure of the
hidden local symmetry cannot be seen in this vacuum.

We need to look for other vacua to realize the hidden local symmetry. A
good candidate can easily be found by looking at the quantum numbers in
Table~\ref{tab:mag}. It is the $q = \bar q \neq 0$ direction that breaks
chiral symmetry down to the diagonal subgroup and gives a mass for the
gauge boson while preserving the baryon number.
We now try to find such a minimum by including soft terms in the
potential.

\subsection{Soft supersymmetry breaking terms in the dual picture}
The mapping of the soft supersymmetry breaking terms into the dual
theory has been studied in Refs.~\cite{Cheng:1998xg, ArkaniHamed:1998wc,
Karch:1998qa, Luty:1999qc, Abel:2011wv}.
In particular, in Ref.~\cite{Luty:1999qc}, simple mapping relations were
found between the electric and the magnetic pictures.
The soft terms in the dual picture can be parametrized by
\begin{eqnarray}
 {\cal L}_{\rm soft}
&=& - \tilde m_q^2 
(|q|^2 + |\bar q|^2 + |q^\prime|^2 + |\bar q^\prime|^2)
- \tilde m_M^2
(|Y|^2 + |Z|^2 + |\bar Z|^2 + |\Phi|^2)
\nonumber \\
&&
- \left(
{m_{\tilde \lambda}\over 2} {\tilde \lambda} {\tilde \lambda}
+
\tilde B m \Lambda Y
+ A h \left(
q^\prime Y \bar q^\prime 
+ q^\prime Z \bar q + q \bar Z \bar q^\prime 
+ q \Phi \bar q
\right)
+ {\rm h.c.}
\right).
\label{eq:soft}
\end{eqnarray}
In the free magnetic range, $N_f \leq N_c /2$, the relations are
\begin{eqnarray}
 \tilde m^2 = 
{2 N_c - N_f \over 3 (N_f + N_c)}
D_R,
\end{eqnarray}
\begin{eqnarray}
 {m_\lambda \over g^2} = - {2 N_c - N_f \over 16 \pi^2}  F_\phi,
\end{eqnarray}
\begin{eqnarray}
 \tilde m_q^2 =
- {N_c - 2 N_f \over 3 (N_f + N_c)} 
 D_R,
\end{eqnarray}
\begin{eqnarray}
 \tilde m_M^2 =
{2 (N_c - 2 N_f) \over 3 (N_f + N_c)} 
 D_R,
\end{eqnarray}
\begin{eqnarray}
 {m_{\tilde \lambda}\over g_H^2} 
= - {2 N_f - N_c \over 16 \pi^2}  F_\phi,
\end{eqnarray}
\begin{eqnarray}
 A = 2 (\gamma_M + \gamma_q + \gamma_{\bar q}) F_\phi,
\end{eqnarray}
where $D_R$ and $F_\phi$ are parameters, and $\gamma_M$, $\gamma_q$ and
$\gamma_{\bar q}$ are anomalous dimensions of $M$, $q$ and $\bar q$,
respectively. The gauge couplings $g$ and $g_H$ are those of the
electric theory ($SU(N_c)$) and the magnetic theory ($SU(N_f)$),
respectively.

We should add positive $\tilde m^2$ in the electric picture. This means
$D_R > 0$. Therefore,
\begin{eqnarray}
 \tilde m_q^2 < 0, \ \ \ \tilde m_M^2 > 0.
\end{eqnarray}
It is interesting to note that $q = q^\prime = \bar q = \bar q^\prime =
0$ is an unstable point. This indeed triggers the chiral symmetry
breaking.

\subsection{Hidden local symmetry in the dual theory}

The scalar fields have potential from three sources, the $F$-term
potential from the superpotential in Eq.~\eqref{eq:superDual}, the
$D$-term from the gauge interactions, and the soft terms in
Eq.~\eqref{eq:soft}:
\begin{eqnarray}
 V = V_F + V_D + V_{\rm soft}.
\end{eqnarray}
The $D$-term potential is given by
\begin{eqnarray}
 V_{D} &=& {g_H^2 \over 2} 
(q^\dagger T^a q - \bar q T^a \bar q^\dagger
+ q^{\prime \dagger} T^a q^\prime 
- \bar q^\prime T^a \bar q^{\prime \dagger} )^2
\nonumber \\
&&
+ {g_{B^\prime}^2 \over 2}
\left(
 {N_c \over N_f} (|q|^2 - |\bar q|^2 )
- {N_f - N_c \over N_f} (|q^\prime|^2 - |\bar q^\prime|^2 )
\right)^2,
\end{eqnarray}
where $g_{B^\prime}$ is the coupling constant of the $U(1)_{B^\prime}$
gauge interaction.

Since $Y$ has a linear term and the positive mass squared in
Eq.~\eqref{eq:soft}, $Y$ can be stabilized at
\begin{eqnarray}
 Y = - {\tilde B m \Lambda \over \tilde m_M^2} .
\end{eqnarray}
Taking $m < B \sim \tilde m$ in the electric picture, the VEV of $Y$ is
large compared to the soft supersymmetry breaking parameters and can be
smaller than $\Lambda$ where the use of duality and weak couplings are
justified. With this VEV, the $q^\prime$ and $\bar q^\prime$ obtain
supersymmetric mass terms and thus decouple.

Once $q^\prime$ and $\bar q^\prime$ get heavy, the system reduces to a
model with $q$, $\bar q$, and $\Phi$. This is the linearized hidden
local symmetry, or the Landau-Ginzburg model in the magnetic picture.
Other light fields, $Y-\langle Y \rangle$, $Z$ and $\bar Z$ are not
directly coupled to this hidden local sector. Since they are stabilized
by the soft terms, one can ignore those fields for the discussion of the
vacuum. The fermionic partners of $Z$ and $\bar Z$ later obtain masses
by the VEV of $q$ and $\bar q$. The fermionic component of $Y$ remains
massless at tree level. The non-perturbative superpotential, $W \propto
(\det M)^{1/N_f}$, together with the VEV of $\Phi$ gives a mass to this
field.

By minimizing the potential, one can find a vacuum with
\begin{eqnarray}
 q = \bar q = v {\bf 1} \neq 0, \ \ \ \Phi = v_\Phi {\bf 1} \neq 0,
\label{eq:vac}
\end{eqnarray}
where the chiral symmetry is spontaneously broken and all the gauge
bosons obtain masses while, of course, the pions remain massless.
In this vacuum, the dual gauge group $SU(N_f)$ and the diagonal part of
the chiral symmetry $SU(N_f)_V \in SU(N_f)_L$ $\times$ $SU(N_f)_R$ is
locked, leaving the global $SU(N_f)_{C+L+R}$ symmetry (the isospin
symmetry) unbroken. This is exactly the structure of the hidden local
symmetry, where $SU(N_f)_{\rm local}$ is now identified as the magnetic
$SU(N_f)$ gauge group.

The pion decay constant $f_\pi$ and the $a$ parameter are respectively
given by
\begin{eqnarray}
 f_\pi^2 = 2 (v^2 + 2 v_\Phi^2),
\end{eqnarray}
and
\begin{eqnarray}
 a = {v^2 \over v^2 + 2 v_\Phi^2}.
\end{eqnarray}
As we know already, $a>1$ cannot be realized in this model. 

\begin{figure}[t]
 \begin{center}
  \includegraphics[width=8cm]{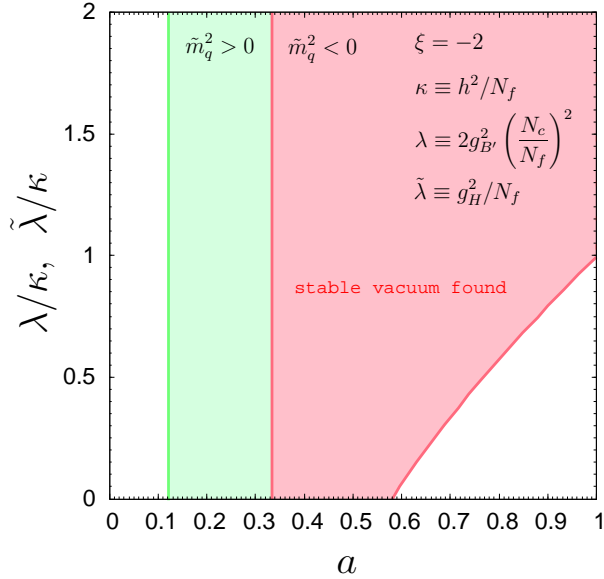}
 \end{center}
\caption{Region where a stable vacuum with chiral symmetry breaking is
found.}  \label{fig:region}
\end{figure}

We show in Figure~\ref{fig:region} the parameter region where a stable
minimum with the VEVs in Eq.~\eqref{eq:vac} is found. In the figure, we
have used a relation $\tilde m_M^2 = - 2 \tilde m_q^2$.
When this relation is imposed, the stability of the vacuum requires,
\begin{eqnarray}
 {\lambda \over \kappa} > {3 a^2 - 1 \over 2 a},\ \ \ 
 {\tilde \lambda \over \kappa} > {3 a^2 - 1 \over 2 a},
\end{eqnarray}
and
\begin{eqnarray}
 3 a (1 + a) - (3 a - 1) \sqrt{
a ( 4 - 3 a)
} > 0.
\label{eq:alow}
\end{eqnarray}
where 
\begin{eqnarray}
 \lambda \equiv 2 g_{B^\prime}^2 \left( N_c \over N_f \right)^2,\ \ \ 
 \tilde \lambda \equiv {g_H^2 \over N_f },\ \ \ 
 \kappa \equiv {h^2 \over N_f}.
\end{eqnarray}
Eq.~\eqref{eq:alow} gives $a > 0.124$.
When $\tilde m_q^2 < 0$ is further imposed, we find a stronger constraint
\begin{eqnarray}
 a > {1 \over 3}.
\end{eqnarray}

The vacuum exists even in the limit where the $U(1)_{B^\prime}$ gauge
interaction is extremely weak. This means that a scale where the
$U(1)_{B^\prime}$ gauge coupling hits the Landau pole can be arbitrarily
high, that we want to be at least higher than the dynamical scale
$\Lambda$ for the model to be well-defined.
Therefore, one can freely take very small mass parameters so that all
the interactions are weak at the scale of $f_\pi$, and thus the analysis
at tree level is reliable.
We now have established the presence of the vacuum in Eq.~\eqref{eq:vac}
in the weakly coupled regime. The vacuum has the same structure of the
hidden local symmetry where the chiral symmetry is spontaneously broken.
The $\rho$ meson is identified as the magnetic gauge boson.
Its mass prevents a magnetic flux of $SU(N_f)$ to enter the vacuum,
describing the confinement in the electric picture if the Seiberg
duality is the electric-magnetic duality.
The fermion components of $q$, $\bar q$, and $\Phi$ all obtain masses
when $a \neq 0, 1$.

Weinberg and Witten have shown that the massless gauge boson cannot be
created by the global current~\cite{Weinberg:1980kq}.
Our $\rho$ meson as the magnetic gauge boson may sound contradicting to
this theorem since the $\rho$ meson couples to the flavor
current. However, the $\rho$ meson in our framework couples to the
global current only after the dual color-flavor locking. Therefore, at
the massless point, the dual color and the flavor symmetry is not
related, and thus the $\rho$ meson does not couple to the global current
in the massless limit.

\section{QCD case ($N_c = 3$, $N_f=2$)}
\label{sec:qcd}

In the following we try to see if how well the vaccum found in the
previous section describes the hadron world in real QCD. Most of the
discussions are qualitative and sometimes speculative, but we can
provide new interpretations of the nature of hadrons.

\subsection{Light mesons as magnetic degrees of freedom}

Since the model, supersymmetric $SU(3)$ gauge theory with $N_f + N_c =
5$ flavors, is in the conformal window, we cannot take the weakly
coupled limit in the infrared.
Nevertheless, one can argue that the dual picture is still more weakly
coupled compared to the electric one. In the regime of the conformal
field theory, anomalous dimensions of $q$, $\bar q$, and $\Phi$ fields
are, respectively, $-1/10$, $-1/10$ and $1/5$, which may be small enough
to justify the expansion in terms of those fields around the free
theory.

In the conformal window, the predictions for the soft parameters are
different from the free magnetic case. In fact, the soft terms (except
for the $\tilde B$ term) approach to zero if the hierarchy of the
dynamical scale and the mass parameters ($m$ and soft terms) is large
enough. However, when we discuss applications to phenomenology, we are
interested in the case with no hierarchy where there is no prediction on
the size or relations among supersymmetry breaking terms. We, therefore,
treat them as free parameters for the analysis.

The boson sector of the model contains various massive and massless
modes. We propose to identify them as light mesons in QCD.
In addition to the massless Nambu-Goldstone particles, $\pi$ (and
$\eta$ at this stage), our model contains massive modes:
\begin{itemize}
 \item $\rho(770)$ as the magnetic gauge boson,
 \item $\omega(782)$ as the $U(1)_{B^\prime}$ gauge boson,
 \item $f_0(600)$, $f_0(980)$, $f_0(1370)$ as the CP-even trace parts of
       $q$,
	$\bar q$ and $\Phi$,
 \item $\eta(1295)$ as the uneaten CP-odd trace part of $q$,
	$\bar q$ and $\Phi$,
 \item $a_0(?)$, $a_0(980)$, $a_0(1450)$ as the CP-even traceless parts of $q$,
	$\bar q$ and $\Phi$, and
 \item $\pi(1300)$ as the uneaten CP-odd traceless part of $q$,
	$\bar q$ and $\Phi$.
\end{itemize}
There should be three CP-even isospin-triplet scalars (called $a_0$) as
mixtures of $q$, $\bar q$ and $\Phi$. We could not find one of them in
the table in \cite{Nakamura:2010zzi}.

In the potential, there are numbers of parameters:
\begin{eqnarray}
 f_\pi^2, \ \ \ a, \ \ \ g_H, \ \ \ g_{B^\prime},\ \ \ 
h,\ \ \ \xi = {\tilde m_M^2 \over \tilde m_q^2}.
\end{eqnarray}
It is probably not very meaningful to try to fit the hadron spectrum
with those parameters since we already know that the phenomenologically
favorable value, $a \sim 2$, cannot be obtained at tree level.
However, from the special shape of the potential, one can find sum rules
independent of parameters:
\begin{eqnarray}
 \sum_{i=1}^3 m_{f_{0,i}}^2 = m_{\eta_H}^2 + m_\omega^2,
\label{eq:sum1}
\end{eqnarray}
\begin{eqnarray}
 \sum_{i=1}^3 m_{a_{0,i}}^2 = m_{\pi_H}^2 + m_\rho^2,
\label{eq:sum2}
\end{eqnarray}
\begin{eqnarray}
 m_{\eta_H}^2 = m_{\pi_H}^2,
\label{eq:sum3}
\end{eqnarray}
where $f_{0,i}$ ($a_{0,i}$) are the three mass eigenstates corresponding
to mixtures of three CP-even isospin singlet (triplet) part of $q$,
$\bar q$, and $\Phi$, and $\eta_H$ and $\pi_H$ are massive states of the
CP-odd isospin singlet and triplet scalars, respectively.
The sum rule in Eq.~\eqref{eq:sum1} can be tested by inputting masses
listed above. The agreement is at the level of $40\%$\footnote{It gets
better when we identify $\eta_H$ as a heavier $\eta$.}, which we think
is good enough since we expect a quantum correction of $O(g_H^2
N_f/(4\pi)^2)$, non-perturbative effects, and also corrections from
finite quark masses which we discuss later.
The second sum rule in Eq.~\eqref{eq:sum2} cannot be tested since we could
not find three $a_0$'s in the hadron spectrum. The sum rule suggest that
the missing one is very light. 
The third one in Eq.~\eqref{eq:sum3} agrees very well although it is not
surprising from the view point of the quark model.

We have so far ignored the finite masses for the quarks. The effects of
the finite quark mass can be taken into account by adding a mass term $m_Q
Q \bar Q$ in the superpotential in the electric picture. In the magnetic
picture, this mass term induces a linear term of the meson,
\begin{eqnarray}
 W_{\rm mass} = m_Q \Lambda \Phi,
\end{eqnarray}
which explicitly breaks the chiral symmetry. The pions obtain masses
from this term.

\subsection{$\eta$}

At the classical level in the dual theory, $\eta$ (it is $\eta^\prime$
in the three flavor language) is massless, but it can obtain a mass through
non-perturbative superpotential. In a meson direction where the dual
squarks can be integrated out, there is a non-perturbative
superpotential,
\begin{eqnarray}
 W_{\rm dyn} \propto \left(
{\rm det} M
\right)^{1/N_f}.
\end{eqnarray}
Since this term explicitly breaks the anomalous $U(1)$ axial symmetry,
$\eta$ obtains a mass. The size is suppressed by a factor of
\begin{eqnarray}
 e^{-{8 \pi^2 / (g_H^2 N_f)}}
\end{eqnarray}
compared to the scale of the $\rho$ meson mass. This is not quite a
suppression factor when we input $g_H \sim 6$, and thus one can
only make a qualitative argument.

The above superpotential is generated by the non-perturbative effect,
such as the instanton configurations of the $\rho$ meson. In this view,
the $\rho$ meson is responsible for the solution to the $U(1)$ problem.

\subsection{QCD string?}

There is a stable string configuration associated with the spontaneous
$U(1)_{B^\prime}$ breaking by the VEVs of the dual squarks. This type of
string has been discussed in Refs.~\cite{Hanany:2003hp, Auzzi:2003fs,
Gorsky:2007ip, Shifman:2007kd, Eto:2007hf} where its connection to
confinement has been emphasized. The string solution involves the
non-trivial configurations of the $SU(N_f)$ gauge fields as well as
$U(1)_{B^\prime}$. The solution is called a non-abelian vortex string as
it non-trivially transforms under the unbroken $SU(N_f)_{C+L+R}$ isospin
symmetry.
The string carries the non-abelian magnetic flux (which is the color
flux in the electric picture), and thus can be interpreted as the QCD
string. It is amusing that the non-trivial configurations of the $\rho$
and $\omega$ mesons can be interpreted as the QCD string!  The QCD
string is a singularity in the vacuum at which $\rho$ and $\omega$
become massless~\cite{'tHooft:1981ht} and thus allows the color flux to
penetrate.

The QCD string should be unstable in the presence of the quarks. In
order to describe the instability at the length scale of ${\cal
O}(\Lambda_{\rm QCD}^{-1})$, one probably needs a larger framework which
explains the nature of $U(1)_{B^\prime}$ and extra flavors. For example,
the meta-stable string and its connection to color confinement has been
discussed in the softly broken ${\cal N}=2$ theory in
Ref.~\cite{Eto:2006dx} where a $U(1)$ factor originates from a breaking
of the $SU(N)$ gauge group.

\subsection{Constituent quarks?}

\begin{figure}[t]
 \begin{center}
  \includegraphics[width=8cm]{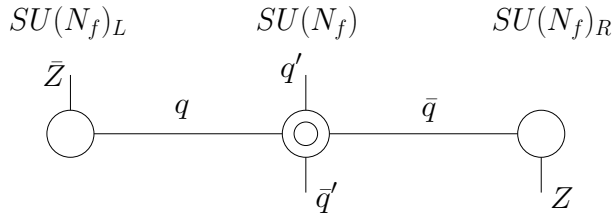}
 \end{center}
\caption{Quiver diagram of the model.}
\label{fig:quiver}
\end{figure}

It is interesting to note that the fermionic components of $Z$, $\bar
Z$, $q^\prime$ and $\bar q^\prime$ have the quantum numbers of quarks if
we gauge the $SU(N_c)_V$ global symmetry. This gauging is a natural
procedure since the global symmetry is an artificially enhanced one by
the introduction of the extra quarks.
The dual quarks $q^\prime$ and $\bar q^\prime$ obtain masses by the VEV
of $Y$. In the language of the hidden local symmetry sketched in
Figure~\ref{fig:quiver}, those are matter fields living in the middle
site.
The fields living in the left and right site, $Z$ and $\bar Z$, obtain
masses through the mixing with $q^\prime$ and $\bar q^\prime$ induced by
the chiral symmetry breaking ($\langle q \rangle = \langle \bar q
\rangle \neq 0$)\footnote{They also obtain masses through the
non-perturbative superpotential.}. These massive ``quarks'' in the dual
picture can naturally be identified as the constituent quarks in the
quark model.

However, these ``quarks'' cannot be the endpoints of the string we
discussed above. The string solution carries the magnetic flux of
$U(1)_{B^\prime}$ whereas the ``quarks'' only have electric
charges\footnote{We thank Naoto Yokoi for discussion on this point.}.
Again, we probably need a larger framework to understand the whole
picture.

\subsection{Baryons}

There are a few ways to describe baryons in this model. 
The simplest one is to introduce a massive vector-like pair of fermions
which transform as fundamental and anti-fundamental under the $SU(N_f)$
dual gauge group (the middle site in Figure~\ref{fig:quiver}). Since
they are massive, one can assume that those are lost in passing through
the Seiberg duality.
In this case, the baryons are flavor singlet and massive in the chiral
symmetric phase, but after the (dual) color-flavor locking by the VEV of
$q$ and $\bar q$, the fermions carry the same quantum numbers as the
proton and the neutron.

Another way is to argue that the skyrmion solution describes the
baryons~\cite{Fujiwara:1984pk}.
In the above two cases, one can explain the universality of the $\rho$
meson couplings, $g_{\rho N N } = g_H$, which was not automatic in the
hidden local symmetry~\cite{Fujiwara:1984pk, Bando:1984pw}.

Of course, the baryons can be three-quark states made of the constituent
quarks we found above. However, in this case, the coupling universality
is not automatic. The coupling depends on the mixing between $q^\prime$
and $Z$ and also on how baryons are composed of.

\section{Application to electroweak symmetry breaking}
\label{sec:ewsb}

The electroweak symmetry breaking in the standard model has the same
structure as chiral symmetry breaking in QCD once we embed the $SU(2)_L$
$\times$ $U(1)_Y$ gauge group into the global symmetry, $SU(N_f)_L$
$\times$ $SU(N_f)_R$. The SQCD model can be thought of as a deformation
of the QCD-like technicolor theory where we have a weakly coupled
description.

For concreteness, we take the model with $N_c = 3$ and $N_f = 2$, and
weakly gauge $SU(N_f)_L$ which is identified as $SU(2)_L$ of the weak
interaction. We embed the $U(1)_Y$ group as $T_3$ generator of the
$SU(N_f)_R$. By this embedding, one can notice that the meson field
$\Phi$ has the same quantum numbers as the Higgs field and its
conjugate, $\Phi = (H, \tilde H)$. The model with $N_c = 3$ and $N_f =
2$ is in the conformal window, and the Higgs field has an anomalous
dimension $1/5$. This model provides us with an almost elementary Higgs
field, and thus phenomenologically friendly~\cite{Fukushima:2010pm}.

The $S$ parameter~\cite{Peskin:1990zt} can be calculated by using the
magnetic picture. There is a tree level contribution to the $S$
parameter from the exchange of the $\rho$ meson. This contribution is
\begin{eqnarray}
 \Delta S = 4\pi \cdot {g_\rho^2 \over m_\rho^4 } 
= {4 \pi \over g_H^2},
\end{eqnarray}
which is independent of the $a$ parameter. If we put $g_H = 6$
as an estimate motivated by QCD, we obtain $\Delta S \sim 0.3$. This is
quite big.
However, given that there is a Higgs boson, the situation is slightly
better compared to the QCD-like technicolor theories if the Higgs boson
is light enough.

\section*{Acknowledgements}

RK thanks Naoto Yokoi, Tadakatsu Sakai, and Zohar Komargodski for useful
discussions, and Yutaka Ookouchi and Mitsutoshi Nakamura for the early
stage of collaboration. RK is supported in part by the Grant-in-Aid for
Scientific Research 21840006 and 23740165 of JSPS.


\begin{thebibliography}{1}
%\cite{Bando:1984ej}
\bibitem{Bando:1984ej}
  M.~Bando, T.~Kugo, S.~Uehara, K.~Yamawaki, T.~Yanagida,
  %``Is rho Meson a Dynamical Gauge Boson of Hidden Local Symmetry?,''
  Phys.\ Rev.\ Lett.\  {\bf 54}, 1215 (1985).

%\cite{Sakai:2004cn}
\bibitem{Sakai:2004cn}
  T.~Sakai, S.~Sugimoto,
  %``Low energy hadron physics in holographic QCD,''
  Prog.\ Theor.\ Phys.\  {\bf 113}, 843-882 (2005).
  [arXiv:hep-th/0412141 [hep-th]].

%\cite{Erlich:2005qh}
\bibitem{Erlich:2005qh}
  J.~Erlich, E.~Katz, D.~T.~Son, M.~A.~Stephanov,
  %``QCD and a holographic model of hadrons,''
  Phys.\ Rev.\ Lett.\  {\bf 95}, 261602 (2005).
  [hep-ph/0501128].

%\cite{Da Rold:2005zs}
\bibitem{Da Rold:2005zs}
  L.~Da Rold, A.~Pomarol,
  %``Chiral symmetry breaking from five dimensional spaces,''
  Nucl.\ Phys.\  {\bf B721}, 79-97 (2005).
  [hep-ph/0501218].

%\cite{Strassler:1997ny}
\bibitem{Strassler:1997ny}
  M.~J.~Strassler,
  %``Messages for QCD from the superworld,''
  Prog.\ Theor.\ Phys.\ Suppl.\  {\bf 131}, 439 (1998)
  [arXiv:hep-lat/9803009].
  %%CITATION = PTPSA,131,439;%%

%\cite{Seiberg:1994rs}
\bibitem{Seiberg:1994rs}
  N.~Seiberg, E.~Witten,
  %``Electric - magnetic duality, monopole condensation, and confinement in N=2 supersymmetric Yang-Mills theory,''
  Nucl.\ Phys.\  {\bf B426}, 19-52 (1994).
  [hep-th/9407087].

%\cite{Seiberg:1994aj}
\bibitem{Seiberg:1994aj}
  N.~Seiberg, E.~Witten,
  %``Monopoles, duality and chiral symmetry breaking in N=2 supersymmetric QCD,''
  Nucl.\ Phys.\  {\bf B431}, 484-550 (1994).
  [hep-th/9408099].

%\cite{'tHooft:1981ht}
\bibitem{'tHooft:1981ht}
  G.~'t Hooft,
  %``Topology of the Gauge Condition and New Confinement Phases in Nonabelian Gauge Theories,''
  Nucl.\ Phys.\  {\bf B190}, 455 (1981).

%\cite{Mandelstam:1974vf}
\bibitem{Mandelstam:1974vf}
  S.~Mandelstam,
  %``Vortices And Quark Confinement In Nonabelian Gauge Theories,''
  Phys.\ Lett.\  {\bf B53}, 476-478 (1975).

%\cite{Carlino:2000ff}
\bibitem{Carlino:2000ff}
  G.~Carlino, K.~Konishi, H.~Murayama,
  %``Dynamics of supersymmetric SU(n(c)) and USp(2n(c)) gauge theories,''
  JHEP {\bf 0002}, 004 (2000).
  [hep-th/0001036];
%\cite{Carlino:2000uk}
%\bibitem{Carlino:2000uk}
  G.~Carlino, K.~Konishi, H.~Murayama,
  %``Dynamical symmetry breaking in supersymmetric SU(n(c)) and USp(2n(c)) gauge theories,''
  Nucl.\ Phys.\  {\bf B590}, 37-122 (2000).
  [hep-th/0005076];
%\cite{Carlino:2001ya}
%\bibitem{Carlino:2001ya}
  G.~Carlino, K.~Konishi, S.~Prem Kumar, H.~Murayama,
  %``Vacuum structure and flavor symmetry breaking in supersymmetric SO(n(c)) gauge theories,''
  Nucl.\ Phys.\  {\bf B608}, 51-102 (2001).
  [hep-th/0104064].

%\cite{Konishi:2005qt}
\bibitem{Konishi:2005qt}
  K.~Konishi, G.~Marmorini, N.~Yokoi,
  %``NonAbelian confinement near nontrivial conformal vacua,''
  Nucl.\ Phys.\  {\bf B741}, 180-198 (2006).
  [hep-th/0511121].

%\cite{Eto:2006dx}
\bibitem{Eto:2006dx}
  M.~Eto, L.~Ferretti, K.~Konishi, G.~Marmorini, M.~Nitta, K.~Ohashi, W.~Vinci, N.~Yokoi,
  %``Non-Abelian duality from vortex moduli: A Dual model of color-confinement,''
  Nucl.\ Phys.\  {\bf B780}, 161-187 (2007).
  [hep-th/0611313].

%\cite{Gorsky:2007ip}
\bibitem{Gorsky:2007ip}
  A.~Gorsky, M.~Shifman, A.~Yung,
  %``N = 1 supersymmetric quantum chromodynamics: How confined non-Abelian monopoles emerge from quark condensation,''
  Phys.\ Rev.\  {\bf D75}, 065032 (2007).
  [hep-th/0701040].

%\cite{Seiberg:1994pq}
\bibitem{Seiberg:1994pq}
  N.~Seiberg,
  %``Electric - magnetic duality in supersymmetric nonAbelian gauge theories,''
  Nucl.\ Phys.\  {\bf B435}, 129-146 (1995).
  [hep-th/9411149].

%\cite{Shifman:2007kd}
\bibitem{Shifman:2007kd}
  M.~Shifman, A.~Yung,
  %``Confinement in N=1 SQCD: One step beyond Seiberg's duality,''
  Phys.\ Rev.\  {\bf D76}, 045005 (2007).
  [arXiv:0705.3811 [hep-th]];
%\cite{Shifman:2011ka}
%\bibitem{Shifman:2011ka}
  M.~Shifman, A.~Yung,
  %``Non-Abelian Duality and Confinement: from N=2 to N=1 Supersymmetric QCD,''
  Phys.\ Rev.\  {\bf D83}, 105021 (2011).
  [arXiv:1103.3471 [hep-th]].

%\cite{Komargodski:2010mc}
\bibitem{Komargodski:2010mc}
  Z.~Komargodski,
  %``Vector Mesons and an Interpretation of Seiberg Duality,''
  JHEP {\bf 1102}, 019 (2011).
  [arXiv:1010.4105 [hep-th]].

%\cite{Harada:1999zj}
\bibitem{Harada:1999zj}
  M.~Harada, K.~Yamawaki,
  %``Conformal phase transition and fate of the hidden local symmetry in large N(f) QCD,''
  Phys.\ Rev.\ Lett.\  {\bf 83}, 3374-3377 (1999).
  [hep-ph/9906445].

%\cite{Aharony:1995zh}
\bibitem{Aharony:1995zh}
  O.~Aharony, J.~Sonnenschein, M.~E.~Peskin, S.~Yankielowicz,
  %``Exotic nonsupersymmetric gauge dynamics from supersymmetric QCD,''
  Phys.\ Rev.\  {\bf D52}, 6157-6174 (1995).
  [hep-th/9507013].

%\cite{Seiberg:1994bz}
\bibitem{Seiberg:1994bz}
  N.~Seiberg,
  %``Exact results on the space of vacua of four-dimensional SUSY gauge theories,''
  Phys.\ Rev.\  {\bf D49}, 6857-6863 (1994).
  [hep-th/9402044].

%\cite{Martin:1998yr}
\bibitem{Martin:1998yr}
  S.~P.~Martin, J.~D.~Wells,
  %``Chiral symmetry breaking and effective Lagrangians for softly broken supersymmetric QCD,''
  Phys.\ Rev.\  {\bf D58}, 115013 (1998).
  [hep-th/9801157].

%\cite{Belyaev:2009ve}
\bibitem{Belyaev:2009ve}
  A.~S.~Belyaev, R.~Sekhar Chivukula, N.~D.~Christensen, H.~-J.~He, M.~Kurachi, E.~H.~Simmons, M.~Tanabashi,
  %``W(L) W(L) Scattering in Higgsless Models: Identifying Better Effective Theories,''
  Phys.\ Rev.\  {\bf D80}, 055022 (2009).
  [arXiv:0907.2662 [hep-ph]].

%\cite{Harada:1992bu}
\bibitem{Harada:1992bu}
  M.~Harada, K.~Yamawaki,
  %``Hidden local symmetry at one loop,''
  Phys.\ Lett.\  {\bf B297}, 151-158 (1992).
  [hep-ph/9210208].

%\cite{Intriligator:2006dd}
\bibitem{Intriligator:2006dd}
  K.~A.~Intriligator, N.~Seiberg, D.~Shih,
  %``Dynamical SUSY breaking in meta-stable vacua,''
  JHEP {\bf 0604}, 021 (2006).
  [hep-th/0602239].

%\cite{Giveon:2008wp}
\bibitem{Giveon:2008wp}
  A.~Giveon, A.~Katz and Z.~Komargodski,
  %``On SQCD with massive and massless flavors,''
  JHEP {\bf 0806}, 003 (2008)
  [arXiv:0804.1805 [hep-th]].
  %%CITATION = JHEPA,0806,003;%%

%\cite{Cheng:1998xg}
\bibitem{Cheng:1998xg}
  H.~-C.~Cheng, Y.~Shadmi,
  %``Duality in the presence of supersymmetry breaking,''
  Nucl.\ Phys.\  {\bf B531}, 125-150 (1998).
  [hep-th/9801146].

%\cite{ArkaniHamed:1998wc}
\bibitem{ArkaniHamed:1998wc}
  N.~Arkani-Hamed, R.~Rattazzi,
  %``Exact results for nonholomorphic masses in softly broken supersymmetric gauge theories,''
  Phys.\ Lett.\  {\bf B454}, 290-296 (1999).
  [hep-th/9804068].

%\cite{Karch:1998qa}
\bibitem{Karch:1998qa}
  A.~Karch, T.~Kobayashi, J.~Kubo, G.~Zoupanos,
  %``Infrared behavior of softly broken SQCD and its dual,''
  Phys.\ Lett.\  {\bf B441}, 235-242 (1998).
  [hep-th/9808178].

%\cite{Luty:1999qc}
\bibitem{Luty:1999qc}
  M.~A.~Luty, R.~Rattazzi,
  %``Soft supersymmetry breaking in deformed moduli spaces, conformal theories, and N=2 Yang-Mills theory,''
  JHEP {\bf 9911}, 001 (1999).
  [hep-th/9908085].

%\cite{Abel:2011wv}
\bibitem{Abel:2011wv}
  S.~Abel, M.~Buican, Z.~Komargodski,
  %``Mapping Anomalous Currents in Supersymmetric Dualities,''
  Phys.\ Rev.\  {\bf D84}, 045005 (2011).
  [arXiv:1105.2885 [hep-th]].

%\cite{Weinberg:1980kq}
\bibitem{Weinberg:1980kq}
  S.~Weinberg, E.~Witten,
  %``Limits on Massless Particles,''
  Phys.\ Lett.\  {\bf B96}, 59 (1980).

%\cite{Nakamura:2010zzi}
\bibitem{Nakamura:2010zzi}
  K.~Nakamura {\it et al.} [ Particle Data Group Collaboration ],
  %``Review of particle physics,''
  J.\ Phys.\ G {\bf G37}, 075021 (2010).
  
%\cite{Hanany:2003hp}
\bibitem{Hanany:2003hp}
  A.~Hanany, D.~Tong,
  %``Vortices, instantons and branes,''
  JHEP {\bf 0307}, 037 (2003).
  [hep-th/0306150].

%\cite{Auzzi:2003fs}
\bibitem{Auzzi:2003fs}
  R.~Auzzi, S.~Bolognesi, J.~Evslin, K.~Konishi, A.~Yung,
  %``NonAbelian superconductors: Vortices and confinement in N=2 SQCD,''
  Nucl.\ Phys.\  {\bf B673}, 187-216 (2003).
  [hep-th/0307287].

%\cite{Eto:2007hf}
\bibitem{Eto:2007hf}
  M.~Eto, K.~Hashimoto, S.~Terashima,
  %``QCD String as Vortex String in Seiberg-Dual Theory,''
  JHEP {\bf 0709}, 036 (2007).
  [arXiv:0706.2005 [hep-th]].

%\cite{Fujiwara:1984pk}
\bibitem{Fujiwara:1984pk}
  T.~Fujiwara, Y.~Igarashi, A.~Kobayashi, H.~Otsu, T.~Sato, S.~Sawada,
  %``AN EFFECTIVE LAGRANGIAN FOR PIONS, rho MESONS AND SKYRMIONS,''
  Prog.\ Theor.\ Phys.\  {\bf 74}, 128 (1985).
  
%\cite{Bando:1984pw}
\bibitem{Bando:1984pw}
  M.~Bando, T.~Kugo, K.~Yamawaki,
  %``Composite Gauge Bosons And 'low-energy Theorems' Of Hidden Local Symmetries,''
  Prog.\ Theor.\ Phys.\  {\bf 73}, 1541 (1985).

%\cite{Fukushima:2010pm}
\bibitem{Fukushima:2010pm}
  H.~Fukushima, R.~Kitano, M.~Yamaguchi,
  %``SuperTopcolor,''
  JHEP {\bf 1101}, 111 (2011).
  [arXiv:1012.5394 [hep-ph]].  

%\cite{Peskin:1990zt}
\bibitem{Peskin:1990zt}
  M.~E.~Peskin, T.~Takeuchi,
  %``A New constraint on a strongly interacting Higgs sector,''
  Phys.\ Rev.\ Lett.\  {\bf 65}, 964-967 (1990);
%\cite{Peskin:1991sw}
%\bibitem{Peskin:1991sw}
%  M.~E.~Peskin, T.~Takeuchi,
  %``Estimation of oblique electroweak corrections,''
  Phys.\ Rev.\  {\bf D46}, 381-409 (1992).
  
\end{thebibliography}
\end{document}